\documentclass{PoS}

\title{On the consistency of the spindown behavior of young and old pulsars}

\ShortTitle{R-mode spindown of pulsars}

\author{Mark G. Alford \\
        Address: Department of Physics, Washington University, St. Louis, Missouri, 63130, USA\\
        E-mail: \email{alford@wuphys.wustl.edu}}

\author{\speaker{Kai Schwenzer}%
       \thanks{We are grateful to Brynmor Haskell, Wynn Ho, Prashant Jaikumar, Feryal \"Ozel and Simin Mahmoodifar for helpful discussions. This research was supported in part by the Offices of Nuclear Physics and High Energy Physics of the U.S. Department of Energy under contracts \#DE-FG02-91ER40628, \#DE-FG02-05ER41375.}
\\
       Address: Department of Physics, Washington University, St. Louis, Missouri, 63130, USA\\
       E-mail: \email{schwenzer@physics.wustl.edu}}

\abstract{We study the spindown of pulsars due to gravitational wave emission and show that r-modes in neutron stars provide a quantitative explanation for the observed low rotation frequencies of young pulsars if the r-mode saturation amplitude is sufficiently large. In contrast for such large saturation amplitudes old hadronic millisecond pulsars would spin down much faster than observed. We discuss resolutions of this apparent discrepancy that could make this mechanism consistent with the observational data.}

\FullConference{Xth Quark Confinement and the Hadron Spectrum\\
		 8Ð12 October 2012\\
		 TUM Campus Garching, Munich, Germany}

\begin{document}

\section{Introduction}
Compact stars are dense enough that they could contain novel phases of dense, strongly-interacting matter in their interior. In particular quark matter presents a challenging possibility \cite{Alford:2007xm}. To probe the physics of the core of a compact star is complicated since all observed electromagnetic radiation originates exclusively from its surface or its magnetosphere and the physics of the crust in between these regions is complicated and poorly constrained. The only direct way to "see inside" the star is to use the same methods by which we know about the interior of the earth or sun, namely to study seismic oscillation modes. In contrast to the latter cases, these oscillations cannot be observed directly in a compact star. Yet, in case they have a non-vanishing quadrupole moment, like for instance r-modes, they couple to gravitational waves that carry away angular momentum and can quickly spin down the star. For pulsars the rotation is directly observed via the pulsed emission which is among the most precise data in physics. Oscillations are damped by dissipative mechanisms, which relates the observable pulsar timing data on stars with large frequencies to the microscopic properties of the matter in the star's interior. 

The observed spin frequencies and spindown rates of pulsars show a striking age dependence. Young stars in particular feature spin frequencies that are more than an order of magnitude below their theoretical Kepler limit. Unstable r-mode oscillations \cite{Andersson:1997xt,Andersson:2000mf} have been suggested as an explanation for this surprising finding. Using semi-analytic expressions that allow to study uncertainties in the microscopic as well as macroscopic description, we have recently shown that r-modes of neutron stars can indeed provide a quantitative explanation for the low frequencies of young pulsars \cite{Alford:2012yn}. In contrast to the final frequency the spindown time depends strongly on the amplitude at which the growth of the unstable oscillation is saturated by non-linear, amplitude-dependent damping processes and to obtain a quick spindown requires large saturation amplitudes.

Old x-ray and radio pulsars, in contrast, are or have been spun up by accretion in binary systems and can reach large rotation frequencies. They are very stable and their observed spindown rates are orders of magnitude smaller than those of young stars. With a saturation amplitude required to explain the low frequencies of young stars, millisecond pulsars would spin out of the instability region on time scales that are significantly smaller than their age. Here we extend our recent analysis \cite{Alford:2012yn} and discuss different scenarios to make the r-mode mechanism for the spindown of young pulsars consistent with the observed data on old pulsars. One possibility is a temperature dependence of the r-mode saturation amplitude, but we show that even a strong temperature dependence does not allow to make the different observations compatible. We propose another scenario based on deconfined quark matter that has the potential to explain the timing data on both young and old pulsars.

\section{R-mode spindown of neutron stars}
In this section we discuss semi-analytic solutions for the spindown evolution of pulsars that are discussed in more detail in \cite{Alford:2012yn}. The basis of this approach is the observation that the microphysical material properties, like the shear and bulk viscosity $\eta$ and $\zeta$, the specific heat $c_V$ and the neutrino emissivity $\epsilon$ that describe a given form of dense matter, depend on the macroscopic parameters relevant for the r-mode evolution, namely the star's temperature $T$ and rotation frequency $\Omega$, over the relevant parameter ranges via power laws
\begin{eqnarray}
\eta=\tilde{\eta}T^{-\sigma}\quad,\quad\zeta\approx \frac{C^{2}\tilde{\Gamma}T^{\delta}}{\omega^{2}}\quad, \quad
c_{V}=\tilde{c}_{V}T^{\upsilon}\quad,\quad\epsilon\approx\tilde{\epsilon}T^{\theta}\:,
\end{eqnarray}
where the subthermal, low temperature limit for the bulk viscosity has been employed which holds for fast spinning neutron stars at $T\ll10^{10}$ K \cite{Alford:2010gw} and the relevant oscillation frequency in the rotating frame is given for an r-mode by $\omega \approx 2/3 \,\Omega$.
In observations pulsars appear as point sources so that only quantities that are averaged over the star are observable. The simple analytic parameter dependence allows to write the macroscopic quantities that are relevant for the r-mode evolution - given by the power radiated in gravitational waves $P_{G}$ , the dissipated power $P_{S}$ and $P_{B}$ due to shear and bulk viscosity, the specific heat of the star $C_{V}$, the total neutrino luminosity $L_{\nu}$ and the moment of inertia $I$ - in a way that the complete microphysics as well as the considered star configuration is encoded in corresponding dimensionless constants $\tilde J$, $\tilde S$, $\tilde V$, $\tilde C_{V}$, $\tilde L$ and $\tilde I$ (see \cite{Alford:2012yn} for more details). Qualitative different forms of matter can have very different power law exponents. For a standard neutron star the shear viscosity is dominated by leptonic scattering processes \cite{Shternin:2008es} with (negative) power law exponent $\sigma\!=\!5/3$ and the neutrino emissivity is due to modified Urca reactions \cite{Friman:1978zq} with exponent $\theta\!=\!8$.

The star evolution for the temperature $T$, the rotation frequency $\Omega$ as well as the dimensionless global r-mode amplitude $\alpha$ are obtained from general energy and angular momentum conservation laws \cite{Owen:1998xg, Ho:1999fh}. 
An r-mode is unstable to the emission of gravitational waves when the growth is faster than the viscous damping $1/\tau_G>1/\tau_V+1/\tau_S$, where $\tau_i=P_i/(2E_m)$ are the corresponding characteristic time scales given in terms of the mode energy $E_m$. Then the amplitude grows exponentially on time scales of the order of seconds and this growth is eventually stopped by a non-linear saturation mechanism. There are several proposed mechanisms \cite{Arras:2002dw, Bondarescu:2008qx,Lindblom:2000az,Alford:2011pi}, but at this point it is not settled which mechanism dominates. We will merely assume that such a mechanism exists that saturates the mode at a finite amplitude $\alpha_{\rm sat}$. In general the saturation amplitude is a function of temperature and frequency which we will model here by a simple power law dependence
\begin{equation}
\alpha_{{\rm sat}}(T,\Omega)=\hat{\alpha}_{{\rm sat}}T^{\beta}\Omega^{\gamma}\:.
 \end{equation}
At saturation and for physical amplitudes $\alpha_{\rm sat}<1$ the residual evolution equations simplify to 
\begin{equation}
\frac{d\Omega}{dt} \approx -\frac{2Q\alpha_{{\rm sat}}^{2}\Omega}{\left|\tau_{G}\right|} \; , \qquad
\frac{dT}{dt}  \approx -\frac{1}{C_{V}}\left(L_{\nu}\!+\! P_{G}\right)\:. \label{eq:evolution}
 \end{equation}
Note that the required dissipation to saturate the r-mode, which is entirely determined by gravitational physics, eventually ends up as heat and appears on the right hand side of the thermal evolution equation.

In \cite{Alford:2012yn} it is shown that in general the thermal evolution of a compact star within the r-mode instability region is always faster than the r-mode spindown. The spindown of a compact star follows thereby a steady-state curve where heating and cooling exactly balance. At temperatures that are high enough that neutrino cooling from the core dominates photon emission from the surface this curve is given for the dominant $m=2$ r-mode and a general form of dense matter by
\begin{equation}
\Omega_{hc}\!\left(T\right)=\left(\frac{3^{8}5^{2}}{2^{15}}\frac{\tilde{L}\Lambda_{{\rm QCD}}^{9-\theta}T^{\theta-2\beta}}{\tilde{J}^{2}\Lambda_{{\rm EW}}^{4}GM^{2}R^{3}\hat{\alpha}_{{\rm sat}}^{2}}\right)^{1/\left(8+2\gamma\right)} \; .\label{eq:steady-state-curve}
\end{equation}

\begin{figure}
\center{\includegraphics[scale=1.]{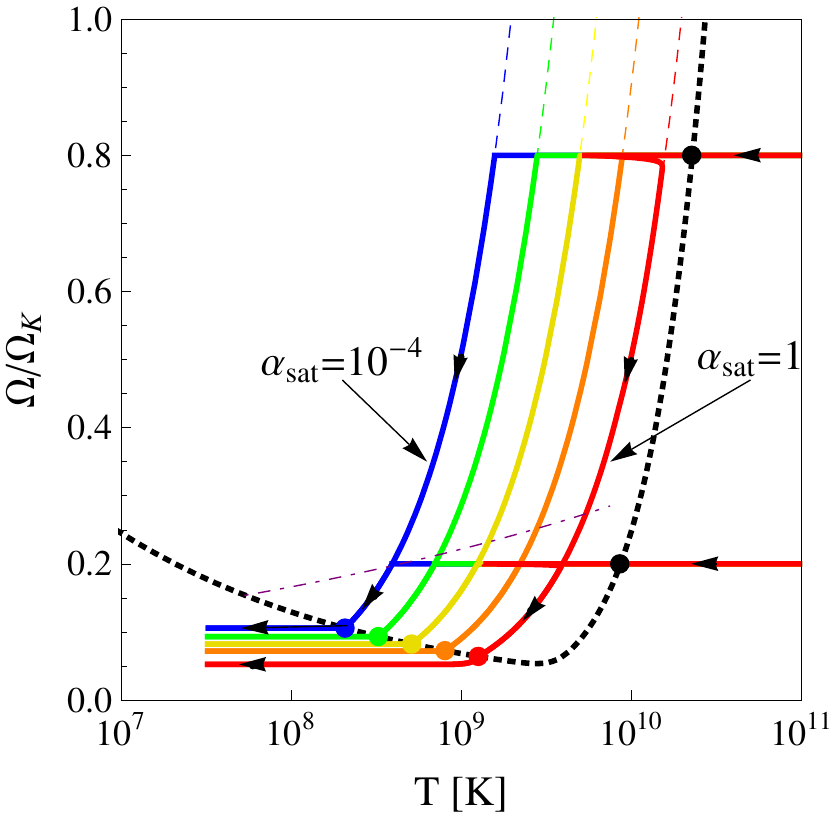}}
\caption{\label{fig:instability-evo}The spindown evolution of a young $1.4\, M_{\odot}$
neutron star with an APR equation of state \cite{Akmal:1998cf} in
temperature-angular velocity space. The boundary of the instability
region of the fundamental $m=2$ r-mode is shown by the dotted curve.
The dashed curves (which are mostly hidden underneath the solid curves)
represent the steady state eq.~(2.4)
where
heating equals cooling and are given for different constant ($\beta\!=\!\gamma\!=\!0$) r-mode amplitudes
ranging from $\alpha_{\rm sat}=10^{-4}$ (left) to $\alpha_{\rm sat}=1$
(right). The dot-dashed curve shows the steady state curve for $\beta=3.5$ and $\alpha_{\rm sat}=10^{-7}$ at the boundary 
(see text). The solid lines show the numerical solution of the evolution
equations for two fiducial initial spin frequencies $\Omega=0.8\,\Omega_{K}$
and $\Omega=0.2\,\Omega_{K}$, where $\Omega_{K}$ is the Kepler frequency. As can be seen, after an initial cooling
phase the evolution simply merges on the appropriate steady state
curve and follows it to the edge of the instability region. The dots
denote the semi-analytic results for the endpoints of the spindown evolution
where the star leaves the instability region, eqs.~(2.5) and (2.6).
}
\end{figure}

A similar semi-analytic expression has been previously given for the low temperature boundary of the instability region \cite{Alford:2010fd}, where shear viscosity dominates the damping. The intersection of the two curves gives then the endpoint of the r-mode spindown evolution, where the star leaves the unstable region and the r-mode is damped. For a standard neutron star $(NS)$ and assuming a constant saturation amplitude ($\beta\!=\!\gamma\!=\!0$) we find
\begin{eqnarray}
T_{f}^{\left(NS\right)}\!\approx\! \left(1.26\cdot10^{9}\,\mathrm{K}\right)\!\!\left(\!\frac{\tilde{S}}{\tilde{S}_{{\rm fid}}}\!\right)^{\!\!3/23}\!\!\!\left(\!\frac{\tilde{L}}{\tilde{L}_{{\rm fid}}}\!\right)^{\!\!-9/92}\!\!\!\left(\!\frac{\tilde{J}}{\tilde{J}_{{\rm fid}}}\!\right)^{\!\!-3/46}\!\!\!\left(\!\frac{M}{1.4\, M_{\odot}}\!\right)^{\!\!-3/46}\!\!\!\left(\!\frac{R}{11.5\,\mathrm{km}}\!\right)^{\!\!-9/92}\!\!\alpha_{{\rm sat}}^{\!9/46} \label{eq:final-temperature-NS}\\
f_{f}^{\left(NS\right)}\!\approx\! \left(61.4\,\mathrm{Hz}\right)\!\!\left(\!\frac{\tilde{S}}{\tilde{S}_{{\rm fid}}}\!\right)^{\!\!3/23}\!\!\!\left(\!\frac{\tilde{L}}{\tilde{L}_{{\rm fid}}}\!\right)^{\!\!5/184}\!\!\!\left(\!\frac{\tilde{J}}{\tilde{J}_{{\rm fid}}}\!\right)^{\!\!-29/92}\!\!\!\left(\!\frac{M}{1.4\, M_{\odot}}\!\right)^{\!\!-29/92}\!\!\!\left(\!\frac{R}{11.5\,\mathrm{km}}\!\right)^{\!\!-87/184}\!\!\alpha_{{\rm sat}}^{\!-5/92} \label{eq:final-frequency-NS}
\end{eqnarray}
where the parameters with the suffix "${\rm fid}$" are the corresponding values for a fiducial $1.4\, M_{\odot}$ neutron star with an APR equation of state \cite{Akmal:1998cf} (see \cite{Alford:2010fd} for more details on the neutron star model). Note the extreme insensitivity of these expressions to the parameters characterizing the poorly known microphysical details like the shear viscosity ($\tilde S$) and the neutrino emissivity ($\tilde L$), as well as to the saturation amplitude $\alpha_{\rm sat}$. Fig. \ref{fig:instability-evo} compares the semi-analytic expressions to a numeric solution for the evolution of a young $1.4\, M_{\odot}$ neutron star for different constant saturation amplitudes ($\beta=\gamma=0$). As can be seen the agreement is very good. The computation of the spindown time requires a solution of the spindown equation eq.~(\ref{eq:evolution}), which gives in case of a neutron star
\begin{equation}
t_{sd}^{\left(NS\right)}\approx\left(12.3\mathrm{\, y}\right)\!\!\left(\!\frac{\tilde{S}}{\tilde{S}_{{\rm fid}}}\!\right)^{\!\!-18/23}\!\!\!\left(\!
\frac{\tilde{L}}{\tilde{L}_{{\rm fid}}}\!\right)^{\!\!-15/92}\!\!\!\left(\!\frac{\tilde{J}}{\tilde{J}_{{\rm fid}}}\!\right)^{\!\!-5/46} \!\!\! \left(\!
\frac{\tilde{I}}{\tilde{I}_{{\rm fid}}}\!\right)\!\!\left(\!\frac{M}{1.4\, M_{\odot}}\!\right)^{\!\!41/46}\!\!\!\left(\!\frac{R}{11.5\,\mathrm{km}}\!
\right)^{\!\!-107/92}\!\!\alpha_{{\rm sat}}^{\!-77/46} \; .\label{eq:spindown-time-NS}
\end{equation}
In contrast to the endpoint of the evolution the spindown time depends strongly on the saturation amplitude. Whereas the spindown is fast for $\alpha_{\rm sat} \gtrsim 10^{-2}$ it becomes longer than the age of observed young pulsars for $\alpha_{\rm sat} \ll 10^{-2}$.

To compare a standard, static r-mode instability analysis to pulsars requires a temperature measurement which is only available for x-ray pulsars that currently accrete. Such an estimate involves many uncertainties since it requires to connect the core temperature to the observed surface temperature and thereby to model the heat transport in the complicated neutron star crust \cite{Chamel:2008LRR}. However, using the semi-analytic results for the spindown evolution it is possible to connect the results of the dynamic r-mode evolution to precise pulsar timing data. This requires the spindown rate eq. (\ref{eq:evolution}) which is given along the spindown curve eq.~(\ref{eq:steady-state-curve}) by
\begin{equation}
\frac{d\Omega}{dt}=12\pi\left(\frac{2^{15}}{3^{8}5^{2}}\right)^{\frac{\theta}{\theta-2\beta}}\left(\frac{\tilde{J}^{2\theta}\Lambda_{EW}^{8\beta}G^{\theta}M^{\theta+2\beta}R^{4\theta-2\beta}\hat{\alpha}_{\rm sat}^{2\theta}\Omega^{7\theta+2\theta\gamma+2\beta}}{\tilde{I}^{\theta-2\beta}\tilde{L}^{2\beta}\Lambda_{QCD}^{2\left(9-\theta\right)\beta}}\right)^{\frac{1}{\theta-2\beta}}\; . \label{eq:spindown-rate}
\end{equation}

\begin{figure}
\center{\includegraphics[scale=1.5]{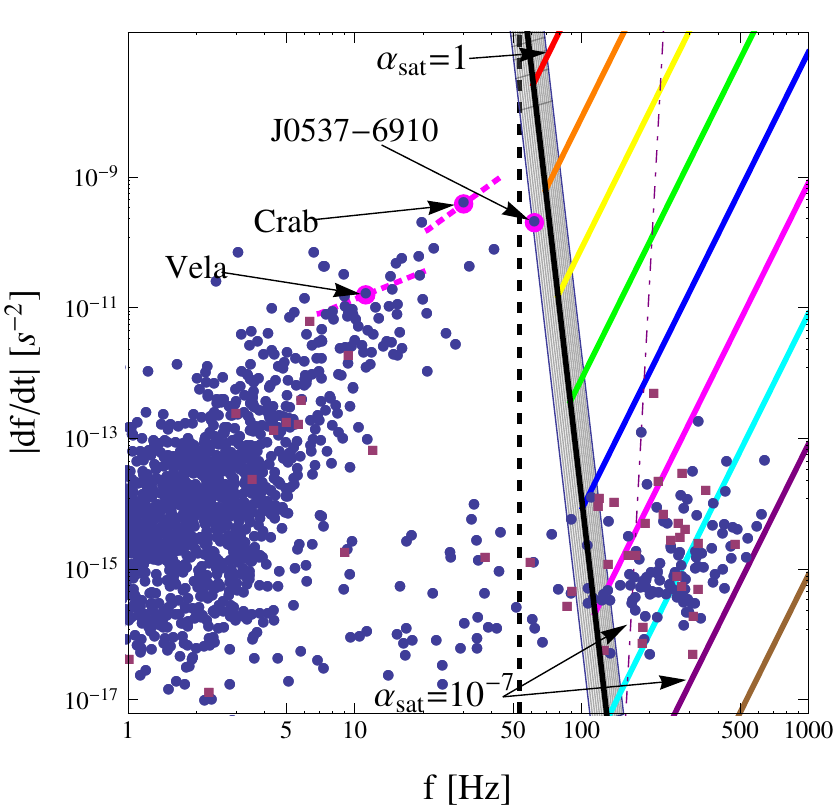}}
\caption{\label{fig:F0F1data}The spindown evolution of a neutron star
compared to observed pulsar data from the ATNF catalog \cite{Manchester:2004bp}. Circles represent stars that spin down whereas squares represent stars that spin up.
The solid lines show the evolution of the $1.4\, M_{\odot}$ star
for different constant saturation amplitudes ($\beta\!=\!\gamma\!=\!0$) ranging from $\alpha_{{\rm sat}}=1$
(top) to $\alpha_{{\rm sat}}=10^{-8}$ (bottom). The dot-dashed line represents the steady-state for $\beta\!=\!3.5$ and $\alpha_{{\rm sat}}=10^{-7}$ at the boundary. The vertical dashed
line to the left shows the lower frequency boundary of the instability
region. The steep solid line is formed by the endpoints of the evolution
for different values of $\alpha$ and thereby represents the boundary
of the instability region in $f$-$\dot{f}$-space, whereas the gray
band reflects the error due to the uncertainty in the underlying parameters. The dotted line segments show the current evolution for two stars
for which a reliable braking index is available \cite{1983bhwd.book.....S,Lyne:1996}. }
\end{figure}

The comparison of the r-mode spindown evolution to pulsar timing data taken from the ATNF catalog \cite{Manchester:2004bp} is shown in fig. \ref{fig:F0F1data}. Here we extend the corresponding plot given in \cite{Alford:2012yn} to include old pulsars which feature much lower spindown rates. The different lines show the r-mode spindown evolution of a $1.4\, M_{\odot}$ neutron star for different saturation amplitudes ranging from $\alpha_{\rm sat}=10^{-8}$ (bottom) to $\alpha_{\rm sat}=1$ (top). Their endpoints map out the boundary of the instability region denoted by the corresponding line with an error band estimated from the uncertainties of the different parameters in eq.~(\ref{eq:final-frequency-NS}) (see \cite{Alford:2012yn} for more details). Let us first consider young stars that have large spindown rates and appear in the upper part of the plot. As can be seen all young stars are likely already outside of the instability region of a neutron star with standard damping mechanisms. Correspondingly, the r-mode mechanism in neutron stars can provide a quantitative explanation of the low spin frequencies of young pulsars. This could be a mere coincidence, but the quantitative agreement with the precise r-mode prediction is quite intriguing. In contrast, stars involving other forms of dense matter with enhanced viscous damping or other dissipative mechanisms in the crust have smaller r-mode instability regions (in a standard T-$\Omega$ diagram) \cite{Haskell:2012} and thereby generally cannot spin down pulsars to similarly low frequencies. 

As seen in fig.~\ref{fig:F0F1data}, most old stars spin at frequencies that are too low for r-modes to be relevant since other mechanisms, like magnetic braking, will spin these stars down to even lower frequencies (stars clustering around a diagonal line in fig.~\ref{fig:F0F1data}). Yet, when a star is in a binary system it can be spun up again by accretion from a low mass companion in a quasi-stable orbit over long time scales and reach nearly kHz frequencies (stars clustering along a horizontal branch in the lower part of fig.~\ref{fig:F0F1data}). These millisecond pulsars are clearly within the instability region of a standard neutron star and their spindown rates are many orders of magnitude lower than those of young stars discussed so far.  Yet, large amplitude r-modes would have spun these stars out of the instability region on time scales much smaller than their current age of up to billions of years. To circumvent this, r-modes in neutron stars would have to saturate at very low (constant) amplitudes $\alpha_{\rm old} < 10^{-7}$ in order to be compatible with the pulsars data. 
Similar bounds have e.g. recently been obtained from x-ray pulsars \cite{Mahmoodifar:2013quw}.
Since pulsars also spin down due to other mechanisms like magnetic dipole emission, those other mechanisms might then explain the scattered data e.g. by different magnetic fields. However, none of the currently proposed saturation mechanisms \cite{Arras:2002dw, Bondarescu:2008qx,Lindblom:2000az,Alford:2011pi} can saturate r-modes at such low saturation amplitudes.

\section{Compatibility of the spin data of young and old stars}
Having seen that the r-mode mechanism in neutron stars can provide a quantitative explanation for the spin frequencies of young pulsars for sufficiently large saturation amplitudes, but would have problems to explain the comprehensive data on old pulsars since these would be well within the instability region, it is interesting to study if there is a consistent scenario that can describe both the r-mode spindown of young pulsars and the observed data on old millisecond pulsars.

According to the results for constant saturation amplitudes in fig.~\ref{fig:F0F1data}, a possibility to explain this finding might be a strong temperature dependence of the saturation amplitude $\alpha_{\rm sat}$ so that the spindown rate is large for hot young stars but many orders of magnitude smaller for cold old stars to meet the strict observational bounds. A qualitatively similar temperature dependence seems to be found in mode coupling models \cite{Arras:2002dw, Bondarescu:2008qx} which present probably the most promising candidate for the r-mode saturation. Within such an explanation of the data the r-mode should be unimportant for the spindown evolution of old stars and other spindown mechanisms like magnetic braking would dominate. In order to model such a temperature dependence of the saturation amplitude we need large amplitudes $0.01\lesssim\alpha_{\rm young} \lesssim0.1$ at core temperatures $10^9\,{\rm K} \lesssim T_{\rm young} \lesssim 10^{10}\,{\rm K}$  relevant for young stars and very small values $10^{-8}\lesssim\alpha_{\rm old} \lesssim 10^{-7}$ at core temperatures $5 \cdot 10^7\,{\rm K} \lesssim T_{\rm old}\lesssim 5\cdot 10^8\,{\rm K}$ realized in old stars. I.e. a change $\alpha_{\rm young}/\alpha_{\rm old} \approx 10^6$ over a temperature range $T_{\rm young}/T_{\rm old} \approx 50$ which requires a large exponent $\beta\approx3.5$ over this range. 
As can be seen from eq.~(\ref{eq:steady-state-curve}) the steady-state curve along which the star spins down is thereby drastically changed from $\Omega \sim T$ for a constant saturation amplitude (in case of neutron star) to the much weaker dependence $\Omega \sim T^{1/8}$. Clearly such a strong temperature dependence of $\alpha_{\rm sat}$ would need to be restricted to the above temperature interval and could not extend up to the boundary of the instability region in fig.~\ref{fig:instability-evo} since this would give unphysically large amplitudes $\alpha_{\rm sat}\gg 1$ at temperatures $T \gtrsim 10^{10}\,{\rm K}$. Assuming that the r-mode amplitude is constant above the considered temperature range, the star would quickly spin down along one of the large amplitude curves in fig.~\ref{fig:instability-evo} and then merge into a more flat segment as shown by the dot-dashed curve for $\beta=3.5$ along which it would exit the instability region after long times. Now let us estimate how the corresponding path would look in fig.~\ref{fig:F0F1data}. According to eq.~(\ref{eq:spindown-rate}) the spindown rate would have an extremely strong dependence on the frequency $d\Omega/dt \sim \Omega^{63}$ for such a temperature dependent saturation amplitude leading to a nearly vertical segment shown by the dot-dashed curve in fig.~\ref{fig:F0F1data}. Clearly such a spindown path would be completely inconsistent with the low spindown rates of millisecond pulsars. Note that we assumed here neutrino cooling, but in case photon emission from the surface dominates for some of these sources this would not qualitatively alter the demonstrated effect. In other words r-modes in a neutron star with standard damping mechanisms cannot simultaneously explain both the low spin frequencies of young pulsars and the high spin frequencies of many old x-ray and radio pulsars.

There is a more exotic but also more interesting possibility. During its initial spindown a young neutron star contracts since centrifugal forces become weaker. This increases the density and could at the corresponding critical value transform hadronic matter in its interior into exotic forms of matter like hyperons or some form of quark matter, whereby a hybrid star is formed. Such forms of matter have a very different bulk viscosity due to non-leptonic strangeness changing reactions with a resonant maximum at significantly lower temperature \cite{Alford:2010gw}, imposing a stability window where the star is stable against gravitational wave emission by r-modes up to large frequencies \cite{Alford:2010fd}. However, in general the subsequent spin-up phase of recycled pulsars in binaries would reverse the process (transforming the star back to neutron matter) and could therefore not simultaneously explain both the spindown of young and old stars. Yet, when the strange matter hypothesis holds and strange quark matter is the true ground state of matter \cite{Witten:1984rs}, once a quark core is formed it would be absolutely stable and could not be transformed back. In fact it would even grow and transform most of the star into strange matter. As has recently been shown \cite{Schwenzer:2012ga}, including important long-range strong interaction corrections to the bulk viscosity, the observed x-ray pulsars are outside of the instability region of a compact star that contains quark matter. Correspondingly, such a quark star could explain the large spin frequencies of observed x-ray pulsars, and as will be discussed in a forthcoming work it also has the potential to explain the data on radio pulsars. Neutron stars that are transformed into quark stars at a later stage of their evolution could thereby provide a universal explanation of the observed pulsar timing data.

\bibliographystyle{h-physrev}
\bibliography{cs}


\end{document}